\documentclass[twocolumn,showpacs,superscriptaddress,prb]{revtex4-1}
\usepackage{graphicx}
\usepackage{color}
\usepackage{amsmath,amsthm,amssymb}
\usepackage{braket}
\usepackage[colorlinks,citecolor=blue,linkcolor=red]{hyperref}
\usepackage[normalem]{ulem}

\begin{document}
 
\title{High-Throughput Search and Prediction of Layered 4f-Materials}

\author{Lin Hou}
\affiliation{Theoretical Division, Los Alamos National Laboratory, Los Alamos, New Mexico 87545, USA}

\author{Ying Wai Li}
\affiliation{Computer, Computational, and Statistical Sciences Division, Los Alamos National Laboratory, Los Alamos, New Mexico 87545, USA}

\author{Christopher Lane}
\email{laneca@lanl.gov}
\affiliation{Theoretical Division, Los Alamos National Laboratory, Los Alamos, New Mexico 87545, USA}

\date{\today} 
\begin{abstract}
The development of multifunctional devices calls for the discovery of new layered materials with novel electronic properties. f-electron systems naturally host a rich set of competing and intertwining phases owning to the presence of strong spin–orbit coupling, electron-electron interactions, and hybridization between itinerant and local electrons. However, very little attention has been devoted to exploring the f-electron family of compounds for new promising layered material candidates. Here, we identify 295  rare earth compounds from across the lanthanide series of elements that exhibit a spectrum of lattice symmetries and electronic properties. In particular,  we find metallic compounds and insulating systems with a broad range of band gaps spanning 0.1 eV to 5.3 eV, which opens new possibilities in infrared quantum sensors, designer photocatalysts, and tunable transistors. The inclusion of 4f-states in a layered system also suggests the possibility of 2D confined heavy-fermion superconductivity and topological semimetals. Our study serves as a springboard to further systematic theoretical investigation of correlation-driven properties of the 4f and other 2D materials composed of heavy elements.
\end{abstract}


\maketitle 

\section{Introduction}
For the past 50 years modern microelectronics, signal processing devices, and data storage technologies have relied on simple metals and semiconductors to provide an exponential growth in processing power that has enabled exceptional leaps in fundamental science, engineering, and communications. However, in the last 15 years computational power has begun to saturate and, with it, Moore’s Law is leveling off\cite{shalf2020future}. The choke point limiting data bandwidth, and ultimately high-throughput information processing, is the limited electronic properties of the underlying conventional materials of the integrated circuits\cite{havemann2001high,murray2018basic}. To overcome these intrinsic limitations and enable new complex integrated devices, new materials hosting rich multifunctional properties are required.

Layered materials are solids with highly anisotropic bonding, i.e. strong covalent bonds within the layers and weak van der Waals type bonds connecting adjacent layers. These systems cover a wide range of novel electronic\cite{kim2015observation}, excitonic\cite{pollmann2015resonant}, valley\cite{rivera2016valley}, topological physics\cite{bansil2016colloquium} that may be tuned to a high degree via doping\cite{lane2019understanding}, electrostatic gating\cite{ye2012superconducting}, and defect engineering\cite{zhou2013intrinsic,jiang2019defect}. Theoretical materials discovery efforts have concentrated on discovering new layered materials composed of various s-, p-, d-block elements, yielding $\sim$5,000 candidate compounds\cite{mounet2018two,cheon2017data,haastrup2018computational,rasmussen2015computational,zhou20192dmatpedia}. So far, several of these material families have been experimentally synthesized and characterized, including graphene, hexagonal boron nitride (hBN), black phosphorus, transition-metal dichalcogenides (TMDs), III-metal monochalcogenides and MXenes\cite{ferrari2015science}.

Compounds composed of heavy elements, such as those in the lanthanide and actinide series, expand the current landscape of layered materials by giving access to phenomena at the intersection of strong spin-orbit coupling, electron-electron correlation effects, and the intertwining of local and itinerant states\cite{witczak2014correlated,hewson1997kondo}. Such systems offer intriguing possibilities for layer-confined unconventional superconductivity, non-trivial topology, and Kondo effects while simultaneously able to be miniaturized and embedded into industrially important semiconductors. A recent set of papers pursues the materialization of layered rare earth compounds by intercalating Eu into a graphite matrix\cite{sokolov20202d} and rare earth doped 2D transition metal dichalcogenides\cite{du2020conversion}, with a handful of cases examining crystalline compounds such as CeI$_2$\cite{kramer2002magnetic}, Ce$_2$Te$_5$\cite{chen2017magnetic}, CeSiI\cite{okuma2021magnetic}, (Y,La,Ce)Te$_3$\cite{ru2006thermodynamic}, Gd(I,Br,Cl,F)$_2$\cite{you2022gadolinium} and YbOCl\cite{yao2019growth}. However, progress in this area has been slow due to a lack of theoretical predictions of new 2D rare earth candidate materials\cite{chen2021recent}.

In this article, we perform a search for all layered rare earth compounds from the major crystallographic databases and elucidate their structural and electronic properties. We identify 295 compounds that span 35 space groups, with each comprised of five unique structure types on average. These materials are mainly composed of reactive nonmetals, metalloids, and a variety of lanthanide elements. Accordingly, a rich array of electronic behaviors are found, covering metallic, insulating, and four possible $Z_2$ topological insulators.

\begin{figure*}[t]
\includegraphics[width=\textwidth]{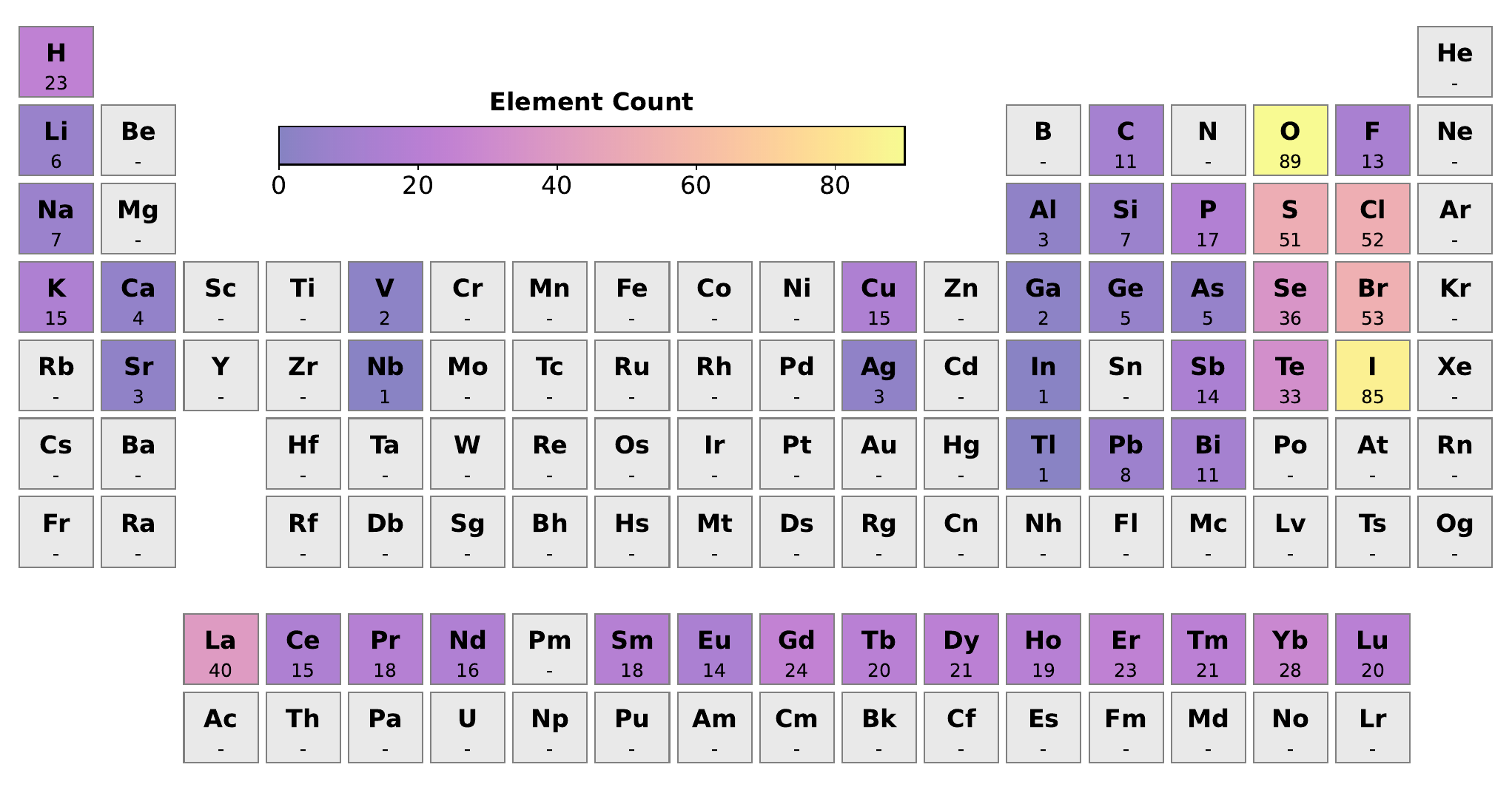}
\caption{(color online) Element distribution of all 295 identified layered 4f compounds plotted within the periodic table. The lighter color indicates a particular element is more prevalent, where each unique elemental species is counted once. }
\label{fig:ptable_distribution}
\end{figure*}

\section{Computational Methods}
\subsection{First-Principles Calculations}
Density functional theory (DFT) calculations were performed using the full potential linearized augmented plane-wave+local orbitals (L/APW+lo) method\cite{blaha1990full,  blaha2001wien2k} as implemented in the WIEN2k code\cite{blaha2020wien2k}. Exchange-correlation effects were treated by using the Perdew-Burke-Ernzerhof generalized gradient approximation (PBE-GGA) \cite{perdew1996generalized}. 1000 (2000) k-points were used to sample the Brillouin zone on a $\Gamma$-centerd grid for unit cell with more than 3 (less than 4) atoms. An energy cutoff of $-6.0$ Ry was used to delineate the core-valence separation. The value of $R_{mt}K_{max}$ was chosen to vary with the size of the smallest atomic species in the unit cell, see Table 1 in the Supplementary Materials for details~\cite{supp}. Furthermore, we set lvns=10 and gmax=25 for materials with H. To gain insight into the sensitivity of the electronic structure to the treatment of relativistic effects and f-electron localization, we consider both non-spin-orbit coupling and finite spin-orbit coupling cases, and f-electrons in the valence and in the core via the open core method.\cite{blaha2020wien2k}

\subsection{Identifying Layered Materials}
To identify layered 4f-materials we must determine the various bonding subunits of a given crystal and their associated dimensionality. This is accomplished by constructing a bond graph and performing a spectral decomposition of the graph Laplacian. Specifically, from the bulk primitive cell of the 3D candidate a $3\times3\times3$ super cell is created. All interatomic distances are then evaluated and chemical bonds are heuristically identified as those for which $d_{ij}<r^{vdW}_{i}+r^{vdW}_{j}-\Delta$, where $d_{ij}$ is the distance between two atoms $i$ and $j$, $r^{vdW}_{i}$ is the van der Waals radius of atom $i$ as determined in Ref.~\onlinecite{alvarez2013cartography} and $\Delta$ is a tolerance factor to account for the experimental variance in $r^{vdW}_{i}$. On average, $\Delta$ is $1.3$ \AA\cite{mounet2018two}. Once all bonds are identified, the adjacency matrix $\mathbf{A}$ is constructed such that $A_{ij}$ equals 1 if there is a bond between atoms $i$ and $j$, otherwise 0. The Laplacian is then $\mathbf{L}=\mathbf{D}-\mathbf{A}$, where $D_{ii}=\sum_{j}A_{ij}$. The number of bonding units in the compound is the dimension of the Null space of $\mathbf{L}$, i.e. the number of zero eigenvalues, with the corresponding eigenvectors specifying the connected units. Finally, the dimensionality of each chemically bonded unit is equal to the rank of the matrix formed by all the vectors connecting a given atomic site to all other equivalent sites in the chemically connected unit in the supercell.

The above procedure was applied to all rare earth compounds from the three major international materials databases: the Inorganic Crystal Structure Database (ICSD), the Crystallography Open Database (COD), and Materials Platform for Data Science (MPDS). To ensure integrity of our search, each material was pre-screened to ensure the structure is crystalline and is not a duplicate crystal structure, similar to Ref.~\onlinecite{hafiz2018high,mounet2018two}.

\begin{figure*}[htbp!]
\centering
\includegraphics[width=\textwidth]{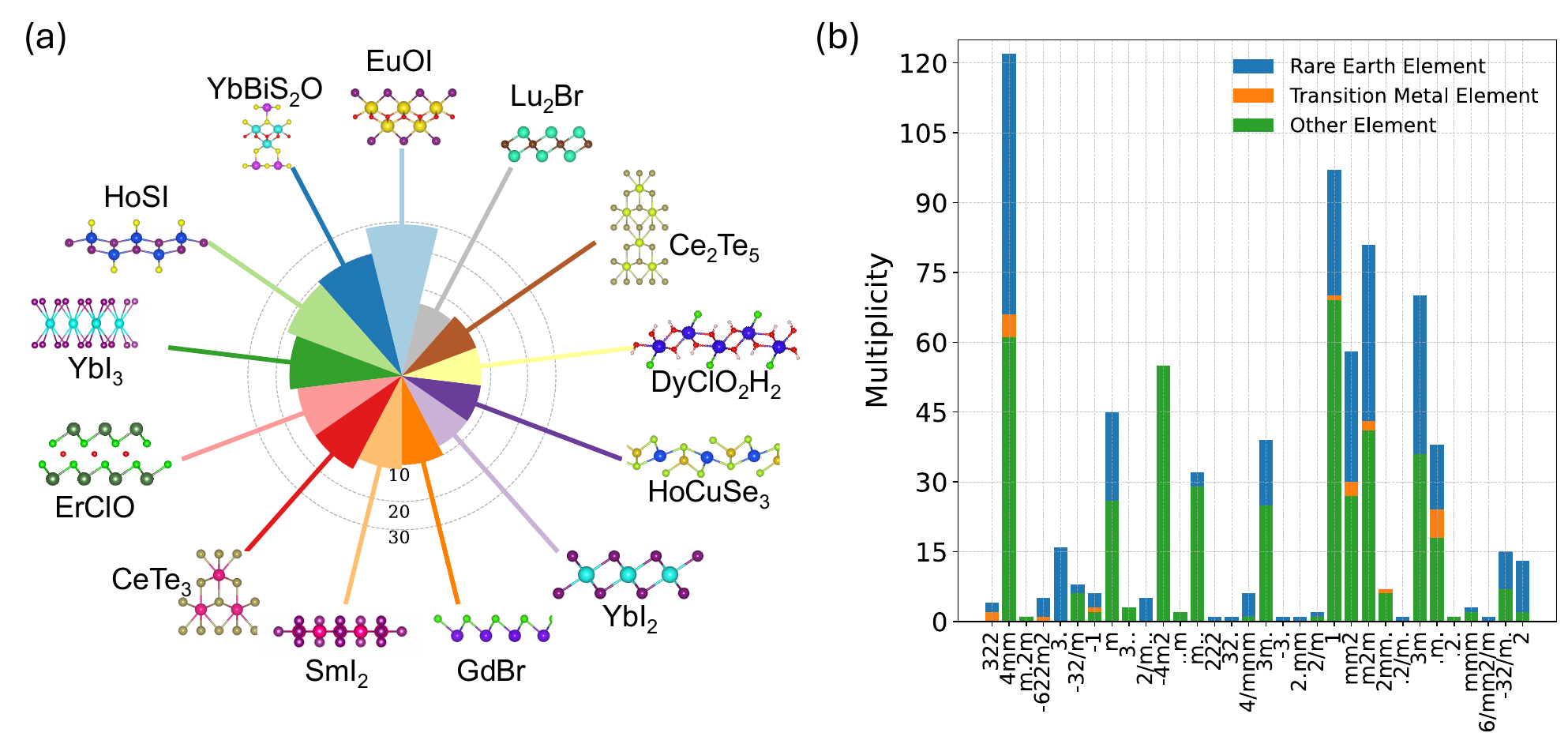}
\caption{(color online) (a) Top 13 Most Frequent layered 4f-material crystal Structure Types. The area of each sector of the pie graph represents the multiplicity of structures. (b) Site symmetry distribution across all 295 layered 4f compounds, categorized by rare earth, transition metal, and other elements. Each unique element site within a material is counted only once. }
\label{fig:sg_distribution}
\end{figure*}

\section{Results and Discussion}
We identify 295 layered 4f compounds sourced from the three major international materials databases: ICSD, COD, and MPDS, all of which are populated by experimental crystal structure data. A complete list of the layered compounds is given in the Supplementary Materials~\cite{supp}. Figure~\ref{fig:ptable_distribution} presents the distribution of the element composition of all identified compounds, where for a given material each unique elemental species was counted once. The 295 compounds are near evenly distributed across the lanthanide series of elements, with the exception of Promethium (Pm), which poses challenges due to its radioactive nature. Notably, near 30\% of all systems feature oxygen or iodine, with significant contributions from the non-metal, other halogen and hydrogen elements. In contrast, only a few transition metals appear in very low frequency. 

Typically, correlated f-electron materials are intermetallic compounds that are composed of elements with 4f or 5f electrons and metals (including post-transition metals and metalloids). These materials typically host three-dimensional (3D) bonding networks and consequently are not layered and hard to cleave. However, if high electronegativity atomic species are introduced, such as oxygen, iodine, chalcogens, other halogens, and other non-metal elements, they are able to occupy interstitial sites and encapsulate the 4f or 5f electrons and metallic elements, hence promoting a layered crystal structure. This process explains the significant presence of oxygen and iodine in the element distribution [Fig.~\ref{fig:ptable_distribution}].   

Though transition metals play a minor role in the identified layered 4f compounds, their inclusion introduces additional electronic degrees of freedom. This expanded space of competing and intertwining interactions further enrich the electronic and magnetic properties of the system. Moreover, they offer opportunities for doping and functionalization, enabling further tailoring of the material's properties for specific applications. However, just a few compounds containing transition-metal ions appear in our search,  suggesting the potential for the synthesis and prediction of new materials.

\subsection{Structure Analysis}
Figure~\ref{fig:sg_distribution}(a) presents the top 13 most frequent crystal structure types of the 295 layered 4f compounds. A full frequency distribution of the 81 total structure types is given in Fig.~1 of the Supplementary Materials~\cite{supp}. The distribution is highly concentrated in a few structure types, with the top 8 structure types representing more than 45\% of crystals. Specifically, structure type EuOI exhibits the highest frequency of occurrence among the compounds, with YbBiS$_2$O, HoSI, and YbI$_3$ structure types following closely in number. These crystal systems cover the tetragonal, orthorhombic, and trigonal crystal systems, respectively, and are consistent with layered systems that typically exhibit large $c/a$ ratios and staggered stacking between layers. All together, these 4 structure types alone account for nearly 30\% of the total number of compounds. The full distribution of structure types for the 295 compounds is given in the Supplementary Materials~\cite{supp}.

\begin{figure*}[ht!]
\includegraphics[width=0.75\textwidth]{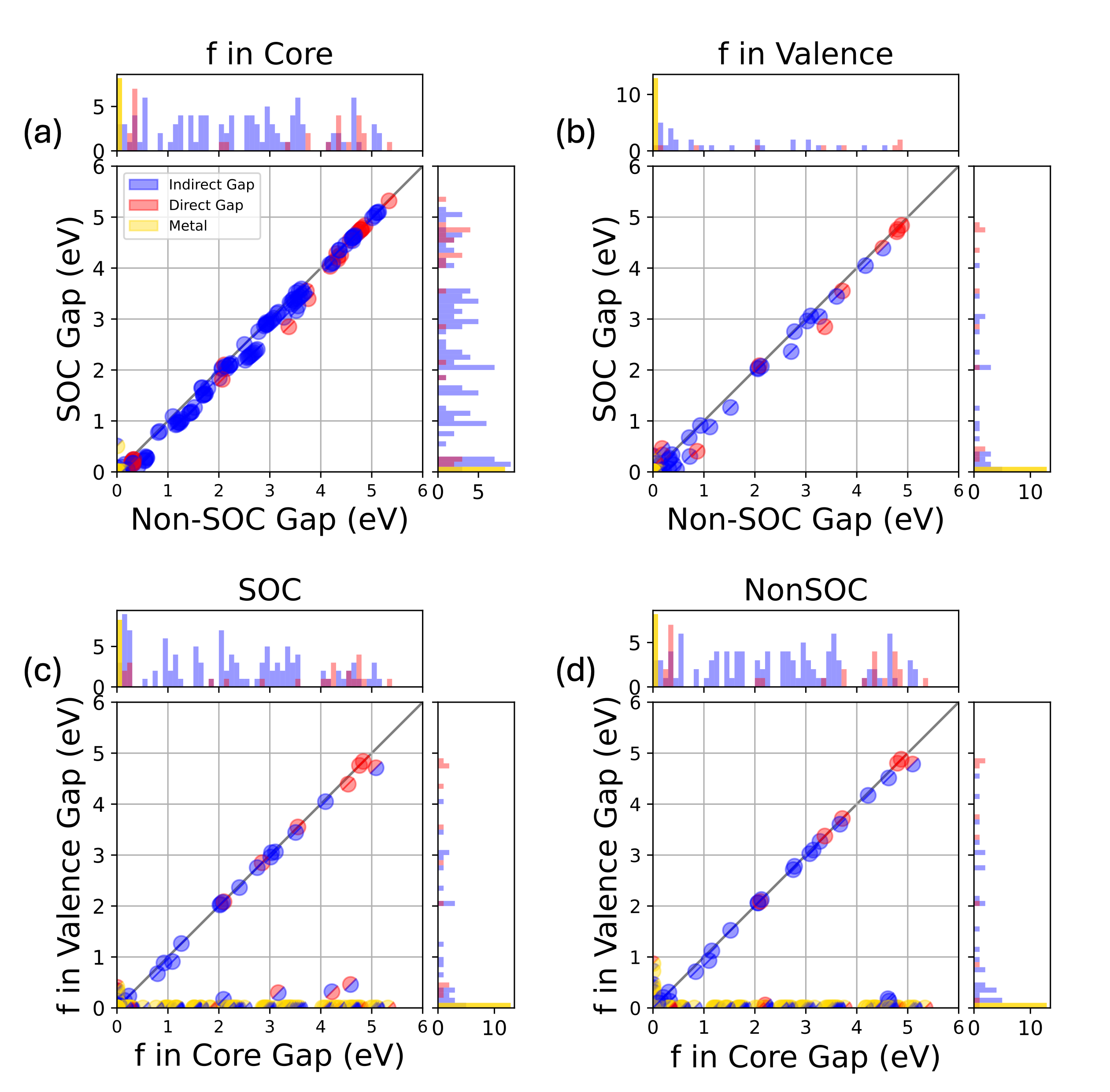}
\caption{(color online) (a) and (b) Comparison of electronic band gaps with and without spin-orbit coupling for f-electrons either located in the core or valence, respectively. (c) and (d) Comparison of electronic band gaps for f-electrons in the core and in the valence for spin-orbit coupling either turned on or off, respectively. Direct, indirect, and zero (metal) band gaps are indicated in red, blue, and gold color, respectively. The zero band gap contribution to the histograms is scaled by a factor of 0.1.} 
\label{fig:gap_distribution}
\end{figure*}

Figure~\ref{fig:sg_distribution} (b) shows the site symmetry distribution for rare earth, transition metal, and the remaining elements across the 295 layered 4f compounds. To facilitate comparison, each unique element site within a material is counted only once. By tabulating site symmetries of the key elements in these crystals, we are able to distinguish the various chemical environment hosted in the aggregate of compounds. Most site symmetries are represented in the 4f compounds, with a notable exception: all cubic point groups are absent. The transition metals are mainly clustered around $.m.$, $4mmm$, and $mm2$ point groups, with the rare earth elements approximately evenly spread across various symmetry groups. The distribution of the remaining elements are concentrated on $1$, 4mm, and -4m2 sites. Interestingly, the most prevalent symmetries found in the broad set of two-dimensional layered materials\cite{mounet2018two}, such as $-3m$, $2/m$, and $-1$, are marginal for the layered 4f compounds. While most of the layered materials studied in the last ten years belong to the hexagonal crystal system, the layered 4f crystals overwhelmingly exhibit tetragonal, orthorhombic, and trigonal lattices, with hexagonal crystals forming a small minority of systems. Additionally, we sought to identify special lattice structures such as the kagome, Lieb, and pyrochlore lattices among the 295 compounds; however, our analysis did not yield any instances of these specific lattice types within our dataset.

A key problem that arises in integrating new materials into existing devices is lattice and chemical matching. Importantly, a wide range of layered and 2D materials, including graphene, transition metal dichalcogenides, hexagonal boron nitride, have demonstrated success in integrating with silicon and other industrially important semiconductors\cite{you2020hybrid,liu2021silicon,cheng20212d}. Since the d-electron ions in these compounds are similarly encapsulated by oxygen, iodine, chalcogens, other halogens, and other non-metal elements, this suggests, our broad set of layered materials may expand the set of integrable layered materials to enrich the properties of current device technologies. 

\subsection{Electronic Properties}
Unlike s-, p-, and d-electrons that can readily diffuse throughout the material,  f-electrons tend to be localized around their atomic center and minimally hybridize with the rest of the system. As a consequence, the f-electron states experience strong correlation effects that can drive a variety of novel phenomena including superconductivity, magnetism, and the Kondo effect. This behavior makes such systems hard to model within standard electronic structure approaches and typically requires higher-level theory to accurately capture their behavior. To provide a baseline of insight into the electronic properties of the 295 layered 4f materials, we performed high-throughput all-electron DFT calculations with the f-levels treated as core and valence electrons, to account for both localized and itinerant f-electron scenarios, respectively. Additionally, to mark the effect of relativistic corrections, calculations were performed with and without spin-orbit coupling.

Figure~\ref{fig:gap_distribution} presents the distribution of electronic band gaps of the 295 layered 4f materials where direct, indirect, and zero (metal) band gaps are indicated with red, blue, and yellow color semicircles. Specifically, in Fig.~\ref{fig:gap_distribution} (a) and (b), we compared the impact of including spin-orbit coupling for f-electrons located in the core and valence, respectively, and in Fig.~\ref{fig:gap_distribution} (c) and (d) compare the effect of open and closed core configurations in the presence of spin-orbit coupling and no relativist effects, respectively.   

In Fig.~\ref{fig:gap_distribution} (a), the inclusion of spin-orbit coupling reduces the band gap by approximately 0.1 eV on average with minimal changes in the band gap type. Interestingly, direct band gap compounds are concentrated near 0.2 eV and 4.8 eV, while those with indirect band gaps are evenly spread between 0.1 eV and 5.2 eV. Despite the removal of the f-electrons from the Fermi level, most materials exhibit metallic behavior regardless of the presence of spin-orbit coupling. Notably, two systems (I$_4$LaSm and Li$_2$P$_2$Pr) gain a finite band gap upon the inclusion of spin-orbit coupling, thereby suggesting these materials to be non-trivial $\mathbb{Z}_2$ topological insulators.

When the f-electrons are included in the valence [Fig.~\ref{fig:gap_distribution} (b)], the effect of spin-orbit coupling is more pronounced for systems with small band gaps ($<$1.5 eV), where some gaps are reduced by up to $~50\%$. This sensitivity at small band gaps stems from the significant number of narrow f-electron bands at the Fermi level. Moreover, the presence of f-electrons in the valence leads to a substantial increase in metallic behavior. The distribution of direct and indirect bands gaps is similar to those in the open core calculations. Notably, four systems (I$_4$LaSm, EuIO, BrEuO and SmIF) gain a finite band gap upon the inclusion of spin-orbit coupling, thereby suggesting these materials to be non-trivial $\mathbb{Z}_2$ topological insulators.

Figure~\ref{fig:gap_distribution} (c) and (d) display very similar features, suggesting that spin-orbit coupling plays a lesser role compared to the treatment of the f-electrons. Many materials transition to being metals, or exhibit an extremely reduced band gap, upon the inclusion of f-states in the valence. Notably, materials containing close/empty shell elements, such as La, experience no change in band gap, with minimal changes in band gap type, as expected. Interestingly, several compounds exhibit an increase in the band gap when the f-electrons are taken out of the core. These systems also display a significant sensitivity to including spin-orbit coupling effects.

A wide range of key applications spanning catalysis, quantum sensors and emitters, and next-generation transistors, rely on the presence of energy band gap in the electronic states. 2D layered materials such as the transition metal dichalcogenides and hexagonal boron nitride have been the workhorse compounds over the past 10-15 years despite their intrinsic electronic band gaps only covering excitation in a 1-2 eV and $\sim 6$ eV window, with almost none otherwise. Here, we find by expanding the family of layered materials to include rare earth elements a range of band gap energies that cover $0.1$ eV to $5.3$ eV. This enlarged range of accessible band gaps opens up new opportunities in narrow band gap (infrared) quantum sensors, designer photocatalysts, tunable transistors and gates. Despite the short comings of bare PBE band gaps~\cite{perdew1983physical,perdew2017understanding}, it provides a baseline for high-throughput material property analysis~\cite{chaves2020bandgap} and a good starting point for future in-depth GW calculations to more accurately capture quasiparticle corrections for a targeted set of applications~\cite{van2015gw}.

The identification of a large number of metallic systems raises the exciting possibility of new broken symmetry phases driven by the coexistence of correlations and strong spin-orbit coupling. In particular, the inclusion of 4f-states in a layered system suggests the possibility of 2D confined heavy-fermion superconductivity, topological semimetals, and strong crystalline anisotropy. Such attractive phases of matter when integrated on a chip opens up new avenues for enhanced mean-free paths, ultra-high conductivity, anisotropic conduction channels. Future detailed many-body perturbation theory calculations are need to elucidate these possibilities. 

\section{Summary}\label{sec:conclusion}
We have expanded the known family of layered materials by identify 295 layered 4f-materials from the three major databases of experimental inorganic crystal structures. Importantly, these materials sit at the virtually unexplored  intersection of strong spin-orbit coupling, electron-electron correlation effects, and the hybridization of local electrons with conduction electrons. Furthermore, their properties expand the phase space of possible materials for quantum sensors and anisotropic conduction channels, necessary for advanced integrated circuits. Our study serves to stimulate new experimental synthesis and characterization, and in-depth theoretical analysis.

\begin{acknowledgments}
This work was carried out under the auspices of the U.S. Department of Energy (DOE) National Nuclear Security Administration under Contract No. 89233218CNA000001. It was supported by the LANL LDRD Program, and in part by the Center for Integrated Nanotechnologies, a DOE BES user facility, in partnership with the LANL Institutional Computing Program for computational resources. Additional computations were performed at the National Energy Research Scientific Computing Center (NERSC), a U.S. Department of Energy Office of Science User Facility located at Lawrence Berkeley National Laboratory, operated under Contract No. DE-AC02-05CH11231 using NERSC award ERCAP0028014. 
\end{acknowledgments}

\bibliography{references}

\end{document}


\title{Supplementary Material: High-Throughput Search and Prediction of \\Layered 4f-Materials}

\author{Lin Hou}
\affiliation{Theoretical Division, Los Alamos National Laboratory, Los Alamos, New Mexico 87545, USA}

\author{Ying Wai Li}
\affiliation{Computer, Computational, and Statistical Sciences Division, Los Alamos National Laboratory, Los Alamos, New Mexico 87545, USA}

\author{Christopher Lane}
\email{laneca@lanl.gov}
\affiliation{Theoretical Division, Los Alamos National Laboratory, Los Alamos, New Mexico 87545, USA}

\date{\today}

\maketitle 

\section{Structure type}
\begin{figure}[ht]
\includegraphics[width=\columnwidth]{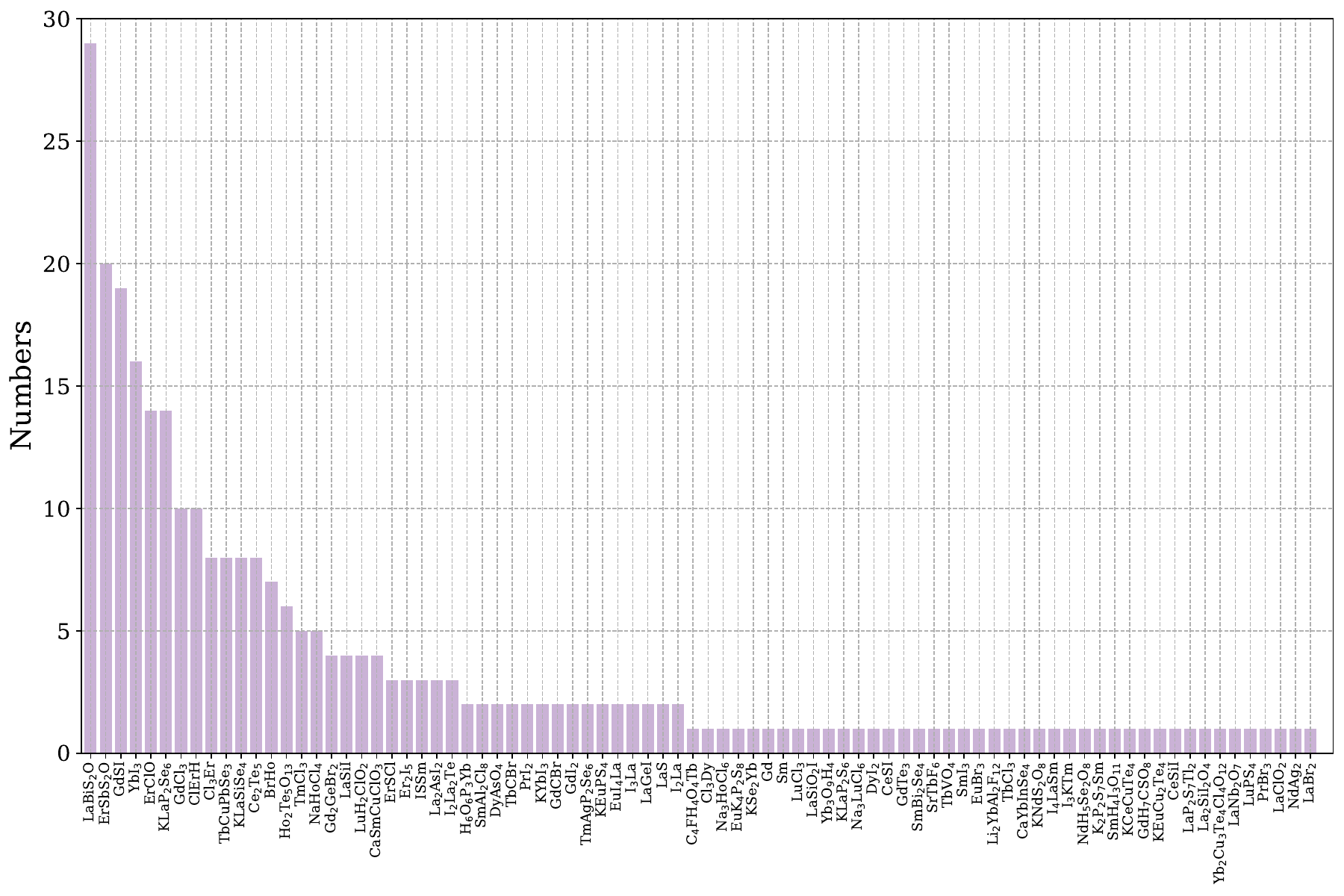}
\caption{295 layered 4f-material crystal Structure Types distribution}
\label{fig:struture_type}
\end{figure}

Figure~\ref{fig:struture_type} presents the full crystal structure type frequency distribution for the 295 4f layered materials. The distribution is highly concentrated in the first few structure types, with tetragonal, orthorhombic, and trigonal crystal systems being the most prevalent.

\clearpage

\section{Computation Details}
The $R_{mt}K_{max}$ used in the self-consistent density functional theory calculations follows the `rule of thumb' given in Table.~\ref{Tab:rkmax}

\begin{table}[ht!]
\label{Tab:rkmax}
\begin{tabular}{ll}
\hline
$R_{mt}K_{max}$ & Smallest Atom In Unit Cell                            \\
\hline
3.0   & H                                        \\
4.5   & Li                                       \\
5.0   & Be, B, Si                                \\
5.5   & C, P                                     \\
6.0   & N, S                                     \\
6.5   & O, Cl, Na, K, Rb, Cs, Mg, Ca, Sr, Ba, Al \\
7.0   & F                                        \\
7.5   & Sc-Cr, Ga-Br, Y-Mo                       \\
8.0   & Mn-Zn, Ru-Cd, In-I, La, Ce, Hf-Re        \\
8.5   & Os-At, Pr-Lu, Ac-Lr        \\
    \hline         
\end{tabular}

\end{table}
\newpage
\section{Data Table}
Table~\ref{Tab:data} presents the chemical, structural, and electronic information for the various layered 4f-materials, along with their database collection IDs. The band gap is given by value and type, i.e. indirect or direct, or zero indicates a metal. If a specific calculation would not converge a dash `-' is used.

\begin{longtable}{rlllclllll}
\caption{Layered 4f Materials Data Table}\\
\label{Tab:data}

ID  & \begin{tabular}[c]{@{}l@{}}Chemical \\ Formula\end{tabular} & Database & \begin{tabular}[c]{@{}l@{}}Database\\ ID\end{tabular} & SG No. & SG Name            & \begin{tabular}[c]{@{}l@{}}Non-SOC\\ f in valence\\ Gap (Type)\end{tabular} & \begin{tabular}[c]{@{}l@{}}SOC\\ f in valence\\ Gap (Type)\end{tabular} & \begin{tabular}[c]{@{}l@{}}Non-SOC\\ f in Core\\ Gap (Type)\end{tabular} & \begin{tabular}[c]{@{}l@{}}SOC\\ f in Core\\ Gap (Type)\end{tabular} \\ \hline

\endfirsthead

\multicolumn{10}{c}{{\tablename\ \thetable{} Layered 4f materials data table continued from previous page}} \\
\toprule
ID  & \begin{tabular}[c]{@{}l@{}}Chemical \\ Formula\end{tabular} & Database & \begin{tabular}[c]{@{}l@{}}Database\\ ID\end{tabular} & SG No. & SG Name            & \begin{tabular}[c]{@{}l@{}}Non-SOC\\ f in valence\\ Gap (Type)\end{tabular} & \begin{tabular}[c]{@{}l@{}}SOC\\ f in valence\\ Gap (Type)\end{tabular} & \begin{tabular}[c]{@{}l@{}}Non-SOC\\ f in Core\\ Gap (Type)\end{tabular} & \begin{tabular}[c]{@{}l@{}}SOC\\ f in Core\\ Gap (Type)\end{tabular} \\ \hline

\endhead

\hline
\multicolumn{10}{c}{{Continued on next page. , I=Indirect Band Gap, D=Direct Band Gap,  '-' = Not Converged}} \\
\endfoot

\hline
\endlastfoot

1 & $ClH_{2}O_{2}Tb$ & COD & 4315693 & 11 & $P2_1/m$ & -- & -- & -- & -- \\
2 & $ClDyH_{2}O_{2}$ & COD & 4315694 & 11 & $P2_1/m$ & -- & -- & -- & -- \\
3 & $PrTe_{3}$ & COD & 4124102 & 40 & $Ama2$ & 0.0 & 0.0 & 0.0 & 0.0 \\
4 & $I_{5}La_{2}$ & COD & 8100408 & 11 & $P2_1/m$ & 0.0 & 0.0 & 0.0 & 0.0 \\
5 & $ILuO$ & COD & 2216822 & 129 & $P4/nmm$ & 3.268 (I) & 3.047 (I) & 3.277 (I) & 3.033 (I) \\
6 & $BrOPr$ & COD & 2232654 & 129 & $P4/nmm$ & 0.0 & 0.0 & 4.337 (D) & 4.174 (D) \\
7 & $NdTe_{3}$ & COD & 4110942 & 63 & $Cmcm$ & 0.0 & 0.0 & 0.0 & 0.0 \\
8 & $Er_{4}I_{5}$ & COD & 1529631 & 12 & $C2/m$ & 0.0 & 0.0 & 0.0 & 0.0 \\
9 & $Cl_{3}Er$ & COD & 6000034 & 12 & $C2/m$ & 0.0 & 0.0 & 4.749 (I) & 4.717 (I) \\
10 & $AlF_{6}LiYb$ & COD & 1519034 & 164 & $P\overline{3}m1$ & 0.0 & 0.0 & 0.0 & 0.0 \\
11 & $ClH_{2}LuO_{2}$ & COD & 4315702 & 62 & $Pnma$ & 4.880 (D) & 4.838 (D) & 4.871 (D) & 4.833 (D) \\
12 & $ILaO$ & COD & 9009171 & 129 & $P4/nmm$ & 3.373 (D) & 2.848 (D) & 3.373 (D) & 2.848 (D) \\
13 & $Cl_{6}HoNa_{3}$ & COD & 8100894 & 14 & $P2_1/c$ & 0.0 & 0.0 & 5.339 (D) & 5.319 (D) \\
14 & $H_{6}O_{6}P_{3}Yb$ & COD & 4322543 & 12 & $C2/m$ & -- & -- & -- & -- \\
15 & $EuKPS_{4}$ & COD & 4319727 & 62 & $Pnma$ & 0.0 & 0.0 & 0.0 & 0.0 \\
16 & $ClH_{2}O_{2}Yb$ & COD & 4315698 & 11 & $P2_1/m$ & 0.0 & 0.0 & 4.788 (D) & 4.750 (D) \\
17 & $BrNdO$ & COD & 9009172 & 129 & $P4/nmm$ & 0.0 & 0.0 & 4.184 (D) & 4.034 (D) \\
18 & $HoTe_{3}$ & COD & 4110939 & 63 & $Cmcm$ & 0.0 & 0.0 & 0.0 & 0.0 \\
19 & $C_{8}H_{12}F_{2}O_{12}Tb_{2}$ & COD & 4329682 & 12 & $C2/m$ & -- & -- & -- & -- \\
20 & $BrLaO$ & COD & 1530049 & 129 & $P4/nmm$ & 3.717 (D) & 3.551 (D) & 3.717 (D) & 3.551 (D) \\
21 & $ClH_{2}HoO_{2}$ & COD & 4315695 & 11 & $P2_1/m$ & -- & -- & -- & -- \\
22 & $IOPr$ & COD & 1530611 & 129 & $P4/nmm$ & 0.0 & 0.0 & 3.753 (D) & 3.398 (I) \\
23 & $IOTm$ & COD & 2310429 & 129 & $P4/nmm$ & 0.0 & 0.0 & 3.553 (I) & 3.256 (I) \\
24 & $Te_{3}Tm$ & COD & 4110941 & 63 & $Cmcm$ & 0.0 & 0.0 & 0.0 & 0.0 \\
25 & $KSe_{2}Yb$ & COD & 8100307 & 166 & $R\overline{3}m$ & 0.0 & 0.0 & 2.008 (I) & 1.838 (I) \\
26 & $Ho_{2}O_{13}Te_{5}$ & COD & 1532259 & 2 & $P\overline{1}$ & 0.0 & 0.0 & 3.173 (I) & 3.135 (I) \\
27 & $ClErH_{2}O_{2}$ & COD & 4315696 & 11 & $P2_1/m$ & 0.0 & 0.0 & 4.721 (D) & 4.687 (D) \\
28 & $ClH_{2}O_{2}Tm$ & COD & 4315700 & 62 & $Pnma$ & 0.0 & 0.0 & 4.826 (D) & 4.790 (D) \\
29 & $TbCl_{3}$ & ICSD & 63543 & 166 & $R\overline{3}m$ & 0.0 & 0.0 & 4.311 (D) & 4.287 (D) \\
30 & $Sm$ & COD & 9010999 & 194 & $P6_3/mmc$ & 0.0 & 0.0 & 0.0 & 0.0 \\
31 & $ClErH$ & COD & 1530725 & 166 & $R\overline{3}m$ & 0.0 & 0.0 & 0.0 & 0.0 \\
32 & $BrOSm$ & COD & 1530050 & 129 & $P4/nmm$ & 0.0 & 0.0 & 4.391 (D) & 4.259 (D) \\
33 & $ErTe_{3}$ & COD & 4110940 & 63 & $Cmcm$ & 0.0 & 0.0 & 0.0 & 0.0 \\
34 & $BrLuO$ & COD & 2214421 & 129 & $P4/nmm$ & 4.171 (I) & 4.048 (I) & 4.223 (I) & 4.093 (I) \\
35 & $I_{2}La$ & COD & 1529708 & 139 & $I4/mmm$ & 0.0 & 0.0 & 0.0 & 0.0 \\
36 & $SmTe_{3}$ & COD & 4110943 & 63 & $Cmcm$ & 0.0 & 0.0 & 0.0 & 0.0 \\
37 & $EuK_{4}P_{2}S_{8}$ & COD & 4319726 & 72 & $Ibam$ & 0.0 & 0.0 & 0.0 & 0.0 \\
38 & $ClH_{2}O_{2}Sm$ & COD & 2012446 & 62 & $Pnma$ & 0.0 & 0.0 & 4.787 (D) & 4.757 (D) \\
39 & $CeH_{2}O_{6}P_{2}$ & COD & 2242273 & 164 & $P\overline{3}m1$ & -- & -- & -- & -- \\
40 & $ClH_{2}O_{2}Yb$ & COD & 4315701 & 62 & $Pnma$ & 0.0 & 0.0 & 4.575 (D) & 4.543 (D) \\
41 & $Gd$ & COD & 1526591 & 47 & $Pmmm$ & -- & -- & -- & -- \\
42 & $ISSm$ & COD & 1008317 & 166 & $R\overline{3}m$ & 0.0 & 0.0 & 1.693 (I) & 1.502 (I) \\
43 & $ClH_{2}LuO_{2}$ & COD & 4315699 & 11 & $P2_1/m$ & 4.802 (D) & 4.759 (D) & 4.799 (D) & 4.760 (D) \\
44 & $ClH_{2}O_{2}Tm$ & COD & 4315697 & 11 & $P2_1/m$ & -- & -- & -- & -- \\
45 & $CeTe_{3}$ & COD & 4124101 & 40 & $Ama2$ & 0.0 & 0.0 & 0.0 & 0.0 \\
46 & $EuKPSe_{4}$ & COD & 4319720 & 62 & $Pnma$ & -- & -- & -- & -- \\
47 & $CCl_{2}Lu_{2}$ & ICSD & 62227 & 166 & $R\overline{3}m$ & 0.930 (I) & 0.910 (I) & 1.099 (I) & 1.087 (I) \\
48 & $ClLa$ & ICSD & 24410 & 166 & $R\overline{3}m$ & 0.0 & 0.0 & 0.0 & 0.0 \\
49 & $CBr_{2}Gd_{2}$ & ICSD & 47226 & 164 & $P\overline{3}m1$ & 0.0 & 0.0 & 0.0 & 0.0 \\
50 & $LaS$ & ICSD & 77831 & 64 & $Cmce$ & 0.0 & 0.0 & 0.0 & 0.0 \\
51 & $Ce_{2}Te_{5}$ & ICSD & 622257 & 63 & $Cmcm$ & 0.0 & 0.0 & 0.0 & 0.0 \\
52 & $EuI_{4}La$ & ICSD & 413999 & 15 & $C2/c$ & 0.0 & 0.0 & 0.0 & 0.0 \\
53 & $CaInSe_{4}Yb$ & ICSD & 67654 & 62 & $Pnma$ & 0.0 & 0.0 & 1.784 (I) & 1.641 (I) \\
54 & $Ho_{2}Te_{5}$ & ICSD & 639768 & 63 & $Cmcm$ & 0.0 & 0.0 & 0.0 & 0.0 \\
55 & $AgP_{2}Se_{6}Tm$ & ICSD & 420304 & 163 & $P\overline{3}1c$ & 0.0 & 0.0 & 1.672 (I) & 1.654 (I) \\
56 & $C_{2}Br_{5}La_{4}$ & ICSD & 418409 & 71 & $Immm$ & -- & -- & -- & -- \\
57 & $H_{8}CeO_{7}P_{3}$ & ICSD & 88339 & 2 & $P\overline{1}$ & -- & -- & -- & -- \\
58 & $Ag_{2}Nd$ & ICSD & 58339 & 191 & $P6/mmm$ & 0.0 & 0.0 & 0.0 & 0.0 \\
59 & $Dy_{2}O_{13}Te_{5}$ & ICSD & 413664 & 2 & $P\overline{1}$ & -- & -- & -- & -- \\
60 & $CaClCuO_{3}Sm$ & ICSD & 86428 & 129 & $P4/nmm$ & 0.0 & 0.0 & 0.0 & 0.0 \\
61 & $Cl_{3}Dy$ & ICSD & 40064 & 63 & $Cmcm$ & 0.0 & 0.0 & 3.620 (I) & 3.590 (I) \\
62 & $I_{4}LaSm$ & ICSD & 413996 & 15 & $C2/c$ & 0.0 & 0.083 (I) & 0.0 & 0.083 (I) \\
63 & $KNdO_{8}S_{2}$ & ICSD & 65620 & 14 & $P2_1/c$ & -- & -- & -- & -- \\
64 & $HClLu$ & ICSD & 62226 & 166 & $R\overline{3}m$ & 0.0 & 0.0 & 0.0 & 0.0 \\
65 & $EuIO$ & ICSD & 27666 & 129 & $P4/nmm$ & 0.0 & 0.299 (D) & 3.518 (I) & 3.166 (I) \\
66 & $Te_{5}Tm_{2}$ & ICSD & 653111 & 63 & $Cmcm$ & 0.0 & 0.0 & 0.0 & 0.0 \\
67 & $CuPbSe_{3}Tb$ & ICSD & 152516 & 62 & $Pnma$ & 0.0 & 0.0 & 1.281 (I) & 1.035 (I) \\
68 & $BiLaOS_{2}$ & ICSD & 252466 & 129 & $P4/nmm$ & 0.0 & 0.0 & 0.0 & 0.0 \\
69 & $As_{2}CeLi_{2}$ & ICSD & 32042 & 164 & $P\overline{3}m1$ & 0.375 (I) & 0.339 (I) & 0.0 & 0.0 \\
70 & $CeI_{4}Sr$ & ICSD & 413997 & 15 & $C2/c$ & 0.0 & 0.0 & 0.097 (I) & 0.095 (I) \\
71 & $AgErP_{2}Se_{6}$ & ICSD & 420303 & 163 & $P\overline{3}1c$ & 0.0 & 0.0 & 1.662 (I) & 1.644 (I) \\
72 & $DyTe_{3}$ & ICSD & 630328 & 63 & $Cmcm$ & 0.0 & 0.0 & 0.0 & 0.0 \\
73 & $Br_{5}Pr_{2}$ & ICSD & 72801 & 11 & $P2_1/m$ & 0.0 & 0.0 & 0.0 & 0.0 \\
74 & $LaSeTe_{2}$ & ICSD & 413173 & 63 & $Cmcm$ & 0.0 & 0.0 & 0.0 & 0.0 \\
75 & $I_{3}KTm$ & ICSD & 72343 & 63 & $Cmcm$ & 0.0 & 0.0 & 0.0 & 0.0 \\
76 & $ClTb$ & ICSD & 23351 & 166 & $R\overline{3}m$ & 0.0 & 0.0 & 0.0 & 0.0 \\
77 & $La_{2}Te_{5}$ & ICSD & 642015 & 63 & $Cmcm$ & 0.0 & 0.0 & 0.0 & 0.0 \\
78 & $C_{2}Br_{5}Ce_{4}$ & ICSD & 418408 & 71 & $Immm$ & 0.0 & 0.0 & 0.0 & 0.0 \\
79 & $CuDyPbSe_{3}$ & ICSD & 152517 & 62 & $Pnma$ & 0.0 & 0.0 & 1.238 (I) & 0.967 (I) \\
80 & $CuPbSe_{3}Yb$ & ICSD & 152521 & 62 & $Pnma$ & 0.0 & 0.0 & 1.210 (I) & 0.966 (I) \\
81 & $I_{2}Tm$ & ICSD & 43731 & 164 & $P\overline{3}m1$ & 0.0 & 0.0 & 0.0 & 0.0 \\
82 & $Al_{2}Cl_{8}Sm$ & ICSD & 56731 & 13 & $P12/a1$ & 0.190 (D) & 0.334 (D) & 0.0 & 0.0 \\
83 & $BiCeOS_{2}$ & ICSD & 80 & 129 & $P4/nmm$ & 0.0 & 0.0 & 0.562 (I) & 0.299 (I) \\
84 & $ClErS$ & ICSD & 21009 & 59 & $Pmmn$ & -- & -- & -- & -- \\
85 & $CeLi_{2}P_{2}$ & ICSD & 42016 & 164 & $P\overline{3}m1$ & 0.327 (I) & 0.265 (I) & 0.0 & 0.0 \\
86 & $Li_{2}P_{2}Pr$ & ICSD & 23259 & 164 & $P\overline{3}m1$ & 0.0 & 0.0 & 0.0 & 0.509 (I) \\
87 & $GeI_{2}La_{2}$ & ICSD & 414171 & 166 & $R\overline{3}m$ & -- & -- & -- & -- \\
88 & $HGdI$ & ICSD & 61035 & 166 & $R\overline{3}m$ & 0.0 & 0.0 & 0.0 & 0.0 \\
89 & $CuLuPbSe_{3}$ & ICSD & 152522 & 62 & $Pnma$ & 1.119 (I) & 0.882 (I) & 1.156 (I) & 0.924 (I) \\
90 & $BrOYb$ & ICSD & 28532 & 129 & $P4/nmm$ & 0.0 & 0.0 & 4.237 (I) & 4.109 (I) \\
91 & $H_{2}ClLaO_{2}$ & ICSD & 6280 & 10 & $P2/m$ & 0.148 (I) & 0.147 (I) & 0.148 (I) & 0.147 (I) \\
92 & $LuTe_{3}$ & ICSD & 642621 & 63 & $Cmcm$ & 0.0 & 0.0 & 0.0 & 0.0 \\
93 & $CuErPbSe_{3}$ & ICSD & 152519 & 62 & $Pnma$ & 0.0 & 0.0 & 1.200 (I) & 0.975 (I) \\
94 & $HLaNb_{2}O_{7}$ & ICSD & 72567 & 123 & $P4/mmm$ & 0.0 & 0.0 & 0.0 & 0.0 \\
95 & $Sm_{2}Te_{5}$ & ICSD & 652655 & 63 & $Cmcm$ & 0.0 & 0.0 & 0.0 & 0.0 \\
96 & $I_{2}Nd$ & ICSD & 72190 & 139 & $I4/mmm$ & 0.0 & 0.0 & 0.0 & 0.0 \\
97 & $BrEuO$ & ICSD & 28531 & 129 & $P4/nmm$ & 0.0 & 0.318 (I) & 4.338 (D) & 4.219 (D) \\
98 & $K_{4}P_{4}S_{14}Sm_{2}$ & ICSD & 161010 & 5 & $C2$ & 0.0 & 0.0 & 0.0 & 0.0 \\
99 & $H_{5}NdO_{8}Se_{2}$ & ICSD & 78051 & 19 & $P2_12_12_1$ & -- & -- & -- & -- \\
100 & $CuPbSe_{3}Yb$ & ICSD & 154613 & 62 & $Pnma$ & 0.0 & 0.0 & 1.186 (I) & 0.937 (I) \\
101 & $CeCuKTe_{4}$ & ICSD & 86654 & 59 & $Pmmn$ & 0.0 & 0.0 & 0.0 & 0.0 \\
102 & $GdIS$ & ICSD & 416462 & 59 & $Pmmn$ & -- & -- & -- & -- \\
103 & $ErISe$ & ICSD & 50194 & 59 & $Pmmn$ & -- & -- & -- & -- \\
104 & $C_{2}H_{7}GdO_{8}S$ & ICSD & 163610 & 2 & $P\overline{1}$ & -- & -- & -- & -- \\
105 & $Dy_{2}Te_{5}$ & ICSD & 630329 & 63 & $Cmcm$ & 0.0 & 0.0 & 0.0 & 0.0 \\
106 & $BrLa$ & ICSD & 23354 & 166 & $R\overline{3}m$ & 0.0 & 0.0 & 0.0 & 0.0 \\
107 & $BrTb$ & ICSD & 23353 & 166 & $R\overline{3}m$ & 0.0 & 0.0 & 0.0 & 0.0 \\
108 & $I_{2}La_{2}Te$ & ICSD & 240698 & 166 & $R\overline{3}m$ & 0.0 & 0.0 & 0.0 & 0.0 \\
109 & $Ge_{2}I_{2}La_{2}$ & ICSD & 59801 & 164 & $P\overline{3}m1$ & 0.0 & 0.0 & 0.0 & 0.0 \\
110 & $NdSeTe_{2}$ & ICSD & 391289 & 63 & $Cmcm$ & 0.0 & 0.0 & 0.0 & 0.0 \\
111 & $I_{3}La$ & ICSD & 31596 & 63 & $Cmcm$ & 1.523 (I) & 1.264 (I) & 1.523 (I) & 1.264 (I) \\
112 & $Br_{2}La$ & ICSD & 65481 & 194 & $P6_3/mmc$ & 0.0 & 0.0 & 0.0 & 0.0 \\
113 & $PrSeTe_{2}$ & ICSD & 412465 & 63 & $Cmcm$ & 0.0 & 0.0 & 0.0 & 0.0 \\
114 & $As_{2}Li_{2}Nd$ & ICSD & 23261 & 164 & $P\overline{3}m1$ & 0.0 & 0.0 & 0.0 & 0.0 \\
115 & $Al_{2}Cl_{8}Eu$ & ICSD & 56734 & 13 & $P12/a1$ & 0.0 & 0.0 & 0.0 & 0.0 \\
116 & $O_{13}Te_{5}Yb_{2}$ & ICSD & 413668 & 2 & $P\overline{1}$ & -- & -- & -- & -- \\
117 & $LaS$ & ICSD & 69581 & 39 & $Abm2$ & 0.0 & 0.0 & 0.0 & 0.0 \\
118 & $I_{2}La_{2}O_{4}Si$ & ICSD & 72654 & 14 & $P2_1/c$ & -- & -- & -- & -- \\
119 & $CuPbSe_{3}Tm$ & ICSD & 152520 & 62 & $Pnma$ & -- & -- & -- & -- \\
120 & $Cu_{2}EuKTe_{4}$ & ICSD & 280193 & 99 & $P4/mm$ & -- & -- & -- & -- \\
121 & $Br_{2}La_{2}P$ & ICSD & 418009 & 164 & $P\overline{3}m1$ & 0.0 & 0.0 & 0.0 & 0.0 \\
122 & $LaTe_{3}$ & ICSD & 642056 & 63 & $Cmcm$ & 0.0 & 0.0 & 0.0 & 0.0 \\
123 & $Ce_{2}I_{2}Si_{2}$ & ICSD & 407246 & 164 & $P\overline{3}m1$ & 0.0 & 0.0 & 0.0 & 0.0 \\
124 & $As_{2}Li_{2}Pr$ & ICSD & 23260 & 164 & $P\overline{3}m1$ & 0.0 & 0.0 & 0.0 & 0.0 \\
125 & $Nd_{2}Te_{5}$ & ICSD & 24181 & 63 & $Cmcm$ & 0.0 & 0.0 & 0.0 & 0.0 \\
126 & $DyIS$ & ICSD & 79107 & 59 & $Pmmn$ & -- & -- & -- & -- \\
127 & $O_{13}Te_{5}Tm_{2}$ & ICSD & 413667 & 2 & $P\overline{1}$ & -- & -- & -- & -- \\
128 & $Er_{2}O_{13}Te_{5}$ & ICSD & 413666 & 2 & $P\overline{1}$ & -- & -- & -- & -- \\
129 & $H_{4}I_{3}O_{11}Sm$ & ICSD & 250551 & 2 & $P\overline{1}$ & -- & -- & -- & -- \\
130 & $I_{2}Yb$ & ICSD & 77907 & 164 & $P\overline{3}m1$ & 0.724 (I) & 0.303 (I) & 0.0 & 0.0 \\
131 & $KLaP_{2}Se_{6}$ & ICSD & 81300 & 14 & $P2_1/c$ & 2.062 (I) & 2.039 (I) & 2.062 (I) & 2.039 (I) \\
132 & $CuHoPbSe_{3}$ & ICSD & 154669 & 62 & $Pnma$ & 0.0 & 0.0 & 1.251 (I) & 0.998 (I) \\
133 & $La_{2}P_{4}S_{14}Tl_{4}$ & ICSD & 195314 & 12 & $C2/m$ & 2.060 (I) & 2.021 (I) & 2.060 (I) & 2.021 (I) \\
134 & $ClTb$ & ICSD & 23352 & 166 & $R\overline{3}m$ & 0.0 & 0.0 & 0.0 & 0.0 \\
135 & $I_{2}La_{2}P$ & ICSD & 418010 & 164 & $P\overline{3}m1$ & 0.0 & 0.0 & 0.0 & 0.0 \\
136 & $Pr_{2}Te_{5}$ & ICSD & 649404 & 63 & $Cmcm$ & 0.0 & 0.0 & 0.0 & 0.0 \\
137 & $TbTe_{3}$ & ICSD & 652952 & 63 & $Cmcm$ & 0.0 & 0.0 & 0.0 & 0.0 \\
138 & $GeI_{2}La_{2}$ & ICSD & 414170 & 164 & $P\overline{3}m1$ & 0.311 (I) & 0.234 (I) & 0.311 (I) & 0.234 (I) \\
139 & $BrDyS$ & ICSD & 79106 & 59 & $Pmmn$ & -- & -- & -- & -- \\
140 & $Br_{5}Ce_{2}$ & ICSD & 167088 & 11 & $P2_1/m$ & 0.0 & 0.0 & 0.0 & 0.0 \\
141 & $HoIS$ & ICSD & 425285 & 59 & $Pmmn$ & -- & -- & -- & -- \\
142 & $ClOYb$ & ICSD & 6077 & 166 & $R\overline{3}m$ & 0.0 & 0.0 & 4.485 (I) & 4.458 (I) \\
143 & $Cl_{4}Cu_{3}O_{12}Te_{4}Yb_{2}$ & ICSD & 419113 & 2 & $P\overline{1}$ & -- & -- & -- & -- \\
144 & $KPrSe_{4}Si$ & ICSD & 416622 & 4 & $P2_1$(No.4) & 0.0 & 0.0 & 0.0 & 0.0 \\
145 & $LuPS_{4}$ & ICSD & 412857 & 14 & $P2_1/c$ & 2.091 (D) & 2.084 (D) & 2.107 (D) & 2.100 (D) \\
146 & $Br_{3}Nd$ & ICSD & 31590 & 63 & $Cmcm$ & 0.0 & 0.0 & 2.241 (I) & 2.128 (I) \\
147 & $Lu_{2}O_{13}Te_{5}$ & ICSD & 413669 & 2 & $P\overline{1}$ & 3.098 (I) & 3.061 (I) & 3.144 (I) & 3.110 (I) \\
148 & $La_{2}SiO_{4}I_{2}$ & MPDS & s1502676 & 2 & $P\overline{1}$ & -- & -- & -- & -- \\
149 & $TbSeI$ & MPDS & s376106 & 59 & $Pmmn$ & -- & -- & -- & -- \\
150 & $ErClO$ & MPDS & s1936405 & 166 & $R\overline{3}m$ & 0.0 & 0.0 & 4.360 (I) & 4.349 (I) \\
151 & $YbSeI$ & MPDS & s376101 & 59 & $Pmmn$ & -- & -- & -- & -- \\
152 & $SmI_{3}$ & MPDS & s1300329 & 63 & $Cmcm$ & 0.0 & 0.0 & 1.430 (I) & 1.154 (I) \\
153 & $ErSbS_{2}O$ & MPDS & s1237956 & 129 & $P4/nmm$ & 0.0 & 0.0 & 0.308 (I) & 0.158 (I) \\
154 & $LaSiI$ & MPDS & s1902101 & 164 & $P\overline{3}m1$ & 0.0 & 0.0 & 0.0 & 0.0 \\
155 & $NaHoCl_{4}$ & MPDS & s1707979 & 2 & $P\overline{1}$ & 0.0 & 0.0 & 4.670 (I) & 4.638 (I) \\
156 & $SrTbF_{6}$ & MPDS & s1827765 & 67 & $Cmme$ & 0.0 & 0.0 & 0.0 & 0.0 \\
157 & $LuSI$ & MPDS & s1937399 & 59 & $Pmmn$ & -- & -- & -- & -- \\
158 & $LaCBr$ & MPDS & s1227665 & 12 & $C2/m$ & -- & -- & -- & -- \\
159 & $PrSI$ & MPDS & s1703862 & 166 & $R\overline{3}m$ & 0.0 & 0.0 & 1.706 (I) & 1.521 (I) \\
160 & $TmCl_{3}$ & MPDS & s1300748 & 12 & $C2/m$ & 0.0 & 0.0 & 5.118 (I) & 5.087 (I) \\
161 & $LuCl_{3}$ & MPDS & s542066 & 167 & $R\overline{3}2/c$ & 4.513 (I) & 4.391 (D) & 4.626 (I) & 4.537 (D) \\
162 & $EuIH$ & MPDS & s1937757 & 129 & $P4/nmm$ & 0.0 & 0.0 & 0.0 & 0.0 \\
163 & $GdBiS_{2}O$ & MPDS & s309739 & 129 & $P4/nmm$ & 0.0 & 0.0 & 0.580 (I) & 0.278 (I) \\
164 & $HoBr$ & MPDS & s542057 & 166 & $R\overline{3}m$ & 0.0 & 0.0 & 0.0 & 0.0 \\
165 & $CaCuEuClO_{3}$ & MPDS & s1900579 & 129 & $P4/nmm$ & 0.0 & 0.0 & 0.0 & 0.0 \\
166 & $DyBrO$ & MPDS & s1903365 & 129 & $P4/nmm$ & 0.0 & 0.0 & 0.0 & 0.0 \\
167 & $PrBiS_{2}O$ & MPDS & s307830 & 129 & $P4/nmm$ & 0.0 & 0.0 & 0.415 (I) & 0.105 (I) \\
168 & $DySbS_{2}O$ & MPDS & s1237954 & 129 & $P4/nmm$ & 0.0 & 0.0 & 0.309 (D) & 0.180 (I) \\
169 & $HoCl_{3}$ & MPDS & s1300746 & 12 & $C2/m$ & 0.0 & 0.0 & 5.053 (I) & 5.024 (I) \\
170 & $GdSI$ & MPDS & s1703873 & 166 & $R\overline{3}m$ & 0.0 & 0.0 & 1.727 (I) & 1.529 (I) \\
171 & $ErBr$ & MPDS & s542062 & 166 & $R\overline{3}m$ & 0.0 & 0.0 & 0.0 & 0.0 \\
172 & $DySI$ & MPDS & s1703875 & 166 & $R\overline{3}m$ & 0.0 & 0.0 & 1.738 (I) & 1.533 (I) \\
173 & $CeBrO$ & MPDS & s1903359 & 129 & $P4/nmm$ & -- & -- & -- & -- \\
174 & $NdBiS_{2}O$ & MPDS & s309736 & 129 & $P4/nmm$ & 0.0 & 0.0 & 0.557 (I) & 0.248 (I) \\
175 & $DyBiS_{2}O$ & MPDS & s309735 & 129 & $P4/nmm$ & 0.0 & 0.0 & 0.541 (I) & 0.215 (I) \\
176 & $HoClO$ & MPDS & s1936402 & 166 & $R\overline{3}m$ & 0.0 & 0.0 & 3.395 (I) & 3.323 (I) \\
177 & $YbI_{3}$ & MPDS & s1300578 & 148 & $R\overline{3}$ & 0.0 & 0.0 & 2.719 (I) & 2.379 (I) \\
178 & $GdSeI$ & MPDS & s376100 & 59 & $Pmmn$ & -- & -- & -- & -- \\
179 & $GdSbS_{2}O$ & MPDS & s1237952 & 129 & $P4/nmm$ & 0.0 & 0.0 & 0.285 (D) & 0.190 (D) \\
180 & $LuI_{3}$ & MPDS & s1300579 & 148 & $R\overline{3}$ & 2.713 (I) & 2.362 (I) & 2.753 (I) & 2.404 (I) \\
181 & $GdCl_{3}$ & MPDS & s1301084 & 63 & $Cmcm$ & 0.0 & 0.0 & 2.928 (I) & 2.902 (I) \\
182 & $ErSBr$ & MPDS & s307276 & 59 & $Pmmn$ & -- & -- & -- & -- \\
183 & $NaDyCl_{4}$ & MPDS & s1707978 & 2 & $P\overline{1}$ & 0.0 & 0.0 & 4.650 (I) & 4.618 (I) \\
184 & $LuSeI$ & MPDS & s376107 & 59 & $Pmmn$ & -- & -- & -- & -- \\
185 & $NdSbS_{2}O$ & MPDS & s307831 & 129 & $P4/nmm$ & 0.0 & 0.0 & 0.330 (D) & 0.243 (D) \\
186 & $HoBr_{3}$ & MPDS & s1300321 & 63 & $Cmcm$ & 0.0 & 0.0 & 2.220 (I) & 2.097 (I) \\
187 & $YbSI$ & MPDS & s1937398 & 59 & $Pmmn$ & -- & -- & -- & -- \\
188 & $Gd_{2}GeBr_{2}$ & MPDS & s1923103 & 164 & $P\overline{3}m1$ & 0.0 & 0.0 & 0.050 (I) & 0.0 \\
189 & $GdI_{2}$ & MPDS & s1300639 & 194 & $P6_3/mmc$ & 0.0 & 0.0 & 0.0 & 0.0 \\
190 & $EuIF$ & MPDS & s1800106 & 129 & $P4/nmm$ & 0.0 & 0.0 & 0.0 & 0.0 \\
191 & $DyAsO_{4}$ & MPDS & s1925690 & 74 & $Imma$ & 0.0 & 0.0 & 2.933 (I) & 2.932 (I) \\
192 & $TmSeF$ & MPDS & s307728 & 166 & $R\overline{3}m$ & 0.0 & 0.0 & 3.004 (I) & 2.949 (I) \\
193 & $ErClO$ & MPDS & s1936404 & 166 & $R\overline{3}m$ & 0.0 & 0.0 & 3.447 (I) & 3.374 (I) \\
194 & $PrI_{3}$ & MPDS & s1300568 & 63 & $Cmcm$ & -- & -- & -- & -- \\
195 & $LaI$ & MPDS & s1301547 & 166 & $R\overline{3}m$ & 0.0 & 0.0 & 0.0 & 0.0 \\
196 & $ErBrO$ & MPDS & s1903357 & 129 & $P4/nmm$ & 0.0 & 0.0 & 3.575 (I) & 3.554 (I) \\
197 & $TbVO_{4}$ & MPDS & s1925692 & 15 & $C2/c$ & 0.0 & 0.0 & 0.0 & 0.0 \\
198 & $LaCI$ & MPDS & s1936154 & 12 & $C2/m$ & -- & -- & -- & -- \\
199 & $GdI_{3}$ & MPDS & s1300318 & 63 & $Cmcm$ & 0.0 & 0.0 & 1.464 (I) & 1.181 (I) \\
200 & $YbCl_{3}$ & MPDS & s1301310 & 63 & $Cmcm$ & 0.0 & 0.0 & 2.924 (I) & 2.892 (I) \\
201 & $YbIF$ & MPDS & s1800112 & 129 & $P4/nmm$ & 0.405 (I) & 0.138 (I) & 0.0 & 0.0 \\
202 & $DyVO_{4}$ & MPDS & s1925691 & 74 & $Imma$ & 0.0 & 0.0 & 2.502 (I) & 2.501 (I) \\
203 & $LuSBr$ & MPDS & s307274 & 59 & $Pmmn$ & -- & -- & -- & -- \\
204 & $TmSeI$ & MPDS & s376102 & 59 & $Pmmn$ & -- & -- & -- & -- \\
205 & $TbBr_{3}$ & MPDS & s1800615 & 12 & $C2/m$ & 0.0 & 0.0 & 4.175 (I) & 4.070 (I) \\
206 & $DyBr_{3}$ & MPDS & s1300320 & 63 & $Cmcm$ & 0.0 & 0.0 & 2.208 (I) & 2.087 (I) \\
207 & $YbI_{3}O_{9}H_{4}O_{2}$ & MPDS & s1711735 & 2 & $P\overline{1}$ & -- & -- & -- & -- \\
208 & $ErIO$ & MPDS & s383717 & 129 & $P4/nmm$ & 0.0 & 0.0 & 0.0 & 0.0 \\
209 & $La_{2}SbI_{2}$ & MPDS & s1713664 & 164 & $P\overline{3}m1$ & 0.0 & 0.0 & 0.0 & 0.0 \\
210 & $TbI_{3}$ & MPDS & s1300319 & 63 & $Cmcm$ & 0.0 & 0.0 & 1.459 (I) & 1.168 (I) \\
211 & $YbSBr$ & MPDS & s307275 & 59 & $Pmmn$ & -- & -- & -- & -- \\
212 & $TmBrF$ & MPDS & s1800108 & 129 & $P4/nmm$ & 0.0 & 0.0 & 0.0 & 0.0 \\
213 & $PrBr_{3}$ & MPDS & s1300387 & 63 & $Cmcm$ & 0.0 & 0.0 & 2.159 (I) & 2.030 (I) \\
214 & $HoSbS_{2}O$ & MPDS & s1237955 & 129 & $P4/nmm$ & 0.0 & 0.0 & 0.307 (D) & 0.179 (I) \\
215 & $GdBrO$ & MPDS & s1903363 & 129 & $P4/nmm$ & 0.0 & 0.0 & 0.0 & 0.0 \\
216 & $GdCBr$ & MPDS & s1703258 & 12 & $C2/m$ & -- & -- & -- & -- \\
217 & $LaBiSe_{2}O$ & MPDS & s1833135 & 129 & $P4/nmm$ & 0.107 (I) & 0.0 & 0.107 (I) & 0.0 \\
218 & $ErI_{3}$ & MPDS & s1300576 & 148 & $R\overline{3}$ & 0.0 & 0.0 & 2.658 (I) & 2.320 (I) \\
219 & $LuSbS_{2}O$ & MPDS & s1237959 & 129 & $P4/nmm$ & 0.132 (I) & 0.0 & 0.332 (I) & 0.171 (I) \\
220 & $ErSeBr$ & MPDS & s1701716 & 59 & $Pmmn$ & -- & -- & -- & -- \\
221 & $TbCBr$ & MPDS & s1702973 & 12 & $C2/m$ & -- & -- & -- & -- \\
222 & $PrI_{2}$ & MPDS & s1300557 & 139 & $I4/mmm$ & 0.0 & 0.0 & 0.0 & 0.0 \\
223 & $GdI_{3}$ & MPDS & s1300572 & 148 & $R\overline{3}$ & 0.0 & 0.0 & 2.560 (I) & 2.240 (I) \\
224 & $Lu_{2}CBr_{2}$ & MPDS & s1708510 & 166 & $R\overline{3}m$ & 0.712 (I) & 0.672 (I) & 0.839 (I) & 0.796 (I) \\
225 & $ErSeF$ & MPDS & s307729 & 166 & $R\overline{3}m$ & 0.0 & 0.0 & 2.969 (I) & 2.917 (I) \\
226 & $TmSI$ & MPDS & s1937397 & 59 & $Pmmn$ & -- & -- & -- & -- \\
227 & $YbBr_{3}$ & MPDS & s1800621 & 148 & $R\overline{3}$ & 0.0 & 0.0 & 3.629 (I) & 3.471 (I) \\
228 & $NaTbCl_{4}$ & MPDS & s1707977 & 2 & $P\overline{1}$ & 0.134 (I) & 0.0 & 4.635 (I) & 4.605 (I) \\
229 & $TmBrO$ & MPDS & s1903367 & 129 & $P4/nmm$ & 0.0 & 0.0 & 0.0 & 0.0 \\
230 & $SmIF$ & MPDS & s1800103 & 129 & $P4/nmm$ & 0.0 & 0.042 (I) & 0.0 & 0.0 \\
231 & $EuBr_{3}$ & MPDS & s1500132 & 63 & $Cmcm$ & 0.058 (D) & 0.169 (I) & 2.203 (I) & 2.092 (I) \\
232 & $HoSeI$ & MPDS & s376104 & 59 & $Pmmn$ & -- & -- & -- & -- \\
233 & $TmBr_{3}$ & MPDS & s1300335 & 148 & $R\overline{3}$ & 0.0 & 0.0 & 3.588 (I) & 3.436 (I) \\
234 & $DyI_{3}$ & MPDS & s1300574 & 148 & $R\overline{3}$ & 0.0 & 0.0 & 2.604 (I) & 2.272 (I) \\
235 & $DyCl_{3}$ & MPDS & s541787 & 12 & $C2/m$ & 0.0 & 0.0 & 5.013 (I) & 4.986 (I) \\
236 & $TbI_{3}$ & MPDS & s1300573 & 148 & $R\overline{3}$ & 0.0 & 0.0 & 2.582 (I) & 2.258 (I) \\
237 & $CaCuGdClO_{3}$ & MPDS & s1900577 & 129 & $P4/nmm$ & 0.0 & 0.0 & 0.0 & 0.0 \\
238 & $YbBiS_{2}O$ & MPDS & s309734 & 129 & $P4/nmm$ & 0.0 & 0.0 & 0.553 (I) & 0.226 (I) \\
239 & $HoSBr$ & MPDS & s307277 & 59 & $Pmmn$ & -- & -- & -- & -- \\
240 & $NdI_{3}$ & MPDS & s1300569 & 63 & $Cmcm$ & 0.0 & 0.0 & 1.420 (I) & 1.151 (I) \\
241 & $NaEuCl_{4}$ & MPDS & s1707975 & 2 & $P\overline{1}$ & 0.180 (I) & 0.460 (D) & 4.612 (I) & 4.584 (I) \\
242 & $TbBrO$ & MPDS & s1903364 & 129 & $P4/nmm$ & 0.0 & 0.0 & 0.0 & 0.0 \\
243 & $PrSbS_{2}O$ & MPDS & s1237948 & 129 & $P4/nmm$ & 0.0 & 0.0 & 0.331 (D) & 0.244 (D) \\
244 & $EuSbS_{2}O$ & MPDS & s1237951 & 129 & $P4/nmm$ & 0.0 & 0.0 & 0.297 (D) & 0.199 (D) \\
245 & $CeI_{2}$ & MPDS & s1721470 & 139 & $I4/mmm$ & 0.0 & 0.0 & 0.0 & 0.0 \\
246 & $Gd_{2}CCl_{2}$ & MPDS & s1400313 & 164 & $P\overline{3}m1$ & 0.0 & 0.0 & 0.812 (I) & 0.772 (I) \\
247 & $TmI_{3}$ & MPDS & s1300577 & 148 & $R\overline{3}$ & 0.0 & 0.0 & 2.691 (I) & 2.344 (I) \\
248 & $ErCl_{3}$ & MPDS & s1301309 & 63 & $Cmcm$ & 0.0 & 0.0 & 2.893 (I) & 2.864 (I) \\
249 & $GdGaI$ & MPDS & s1623208 & 164 & $P\overline{3}m1$ & 0.0 & 0.0 & 0.0 & 0.0 \\
250 & $ErBr_{3}$ & MPDS & s1300334 & 148 & $R\overline{3}$ & 0.0 & 0.0 & 3.549 (I) & 3.402 (I) \\
251 & $TmClO$ & MPDS & s1936406 & 166 & $R\overline{3}m$ & 0.0 & 0.0 & 3.470 (I) & 3.396 (I) \\
252 & $TbSI$ & MPDS & s1937394 & 59 & $Pmmn$ & -- & -- & -- & -- \\
253 & $TmClO$ & MPDS & s1936407 & 166 & $R\overline{3}m$ & 0.0 & 0.0 & 4.365 (I) & 4.353 (I) \\
254 & $HoSeF$ & MPDS & s307730 & 166 & $R\overline{3}m$ & 0.0 & 0.0 & 2.939 (I) & 2.890 (I) \\
255 & $GdTe_{3}$ & MPDS & s525784 & 63 & $Cmcm$ & 0.0 & 0.0 & 0.0 & 0.0 \\
256 & $CeSI$ & MPDS & s1252944 & 61 & $Pbca$ & 0.0 & 0.0 & 2.067 (D) & 1.815 (D) \\
257 & $LuSeF$ & MPDS & s307726 & 166 & $R\overline{3}m$ & 3.027 (I) & 2.961 (I) & 3.077 (I) & 3.017 (I) \\
258 & $ErSI$ & MPDS & s1937396 & 59 & $Pmmn$ & -- & -- & -- & -- \\
259 & $YbBrH$ & MPDS & s1937759 & 129 & $P4/nmm$ & 0.311 (I) & 0.062 (I) & 0.0 & 0.0 \\
260 & $PrI_{2}$ & MPDS & s1300558 & 194 & $P6_3/mmc$ & 0.0 & 0.0 & 0.0 & 0.0 \\
261 & $GdCl$ & MPDS & s542059 & 166 & $R\overline{3}m$ & 0.0 & 0.0 & 0.0 & 0.0 \\
262 & $SmBiS_{2}O$ & MPDS & s309738 & 129 & $P4/nmm$ & 0.0 & 0.0 & 0.587 (I) & 0.276 (I) \\
263 & $KLaP_{2}S_{6}$ & MPDS & s1811367 & 14 & $P2_1/c$ & 2.779 (I) & 2.753 (I) & 2.779 (I) & 2.753 (I) \\
264 & $HoI_{3}$ & MPDS & s1300575 & 148 & $R\overline{3}$ & 0.0 & 0.0 & 2.630 (I) & 2.296 (I) \\
265 & $YbSeF$ & MPDS & s307727 & 166 & $R\overline{3}m$ & 0.0 & 0.0 & 3.037 (I) & 2.979 (I) \\
266 & $HoBrO$ & MPDS & s1903366 & 129 & $P4/nmm$ & 0.0 & 0.0 & 0.0 & 0.0 \\
267 & $La_{2}AsI_{2}$ & MPDS & s1713663 & 164 & $P\overline{3}m1$ & 0.0 & 0.0 & 0.0 & 0.0 \\
268 & $YbSbS_{2}O$ & MPDS & s1237958 & 129 & $P4/nmm$ & 0.0 & 0.0 & 0.310 (D) & 0.175 (I) \\
269 & $SrCuNdClO_{3}$ & MPDS & s1900576 & 129 & $P4/nmm$ & 0.0 & 0.0 & 0.0 & 0.0 \\
270 & $TbSbS_{2}O$ & MPDS & s1237953 & 129 & $P4/nmm$ & 0.0 & 0.0 & 0.310 (I) & 0.175 (I) \\
271 & $DyI_{2}$ & MPDS & s1707656 & 166 & $R\overline{3}m$ & 0.0 & 0.0 & 0.0 & 0.0 \\
272 & $YbBrF$ & MPDS & s1800111 & 129 & $P4/nmm$ & 0.464 (I) & 0.047 (I) & 0.0 & 0.0 \\
273 & $GdBr_{3}$ & MPDS & s1300330 & 148 & $R\overline{3}$ & 0.0 & 0.0 & 3.419 (I) & 3.292 (I) \\
274 & $SmI_{3}$ & MPDS & s1300571 & 148 & $R\overline{3}$ & 0.0 & 0.0 & 2.518 (I) & 2.191 (I) \\
275 & $DyBr_{3}$ & MPDS & s1300332 & 148 & $R\overline{3}$ & 0.0 & 0.0 & 3.487 (I) & 3.347 (I) \\
276 & $SmSbS_{2}O$ & MPDS & s304255 & 129 & $P4/nmm$ & 0.0 & 0.0 & 0.304 (D) & 0.205 (D) \\
277 & $Na_{3}LuCl_{6}$ & MPDS & s1706468 & 14 & $P2_1/c$ & 4.786 (D) & 4.713 (D) & 5.096 (I) & 5.084 (I) \\
278 & $SmIO$ & MPDS & s1401116 & 129 & $P4/nmm$ & 0.0 & 0.0 & 0.0 & 0.0 \\
279 & $TbGaI$ & MPDS & s1623209 & 164 & $P\overline{3}m1$ & 0.0 & 0.0 & 0.0 & 0.0 \\
280 & $NdSI$ & MPDS & s1703863 & 166 & $R\overline{3}m$ & 0.0 & 0.0 & 1.695 (I) & 1.508 (I) \\
281 & $YbCl_{3}$ & MPDS & s1300376 & 12 & $C2/m$ & 0.0 & 0.0 & 5.136 (I) & 5.103 (I) \\
282 & $LaBiS_{2}O$ & MPDS & s1641780 & 11 & $P2_1/m$ & 0.207 (I) & 0.048 (I) & 0.207 (I) & 0.048 (I) \\
283 & $Gd_{2}GeI_{2}$ & MPDS & s1923102 & 166 & $R\overline{3}m$ & 0.0 & 0.0 & 0.116 (I) & 0.0 \\
284 & $KYbI_{3}$ & MPDS & s1710745 & 63 & $Cmcm$ & 0.867 (D) & 0.405 (D) & 0.0 & 0.0 \\
285 & $NaGdCl_{4}$ & MPDS & s1707007 & 2 & $P\overline{1}$ & 0.0 & 0.0 & 4.618 (I) & 4.589 (I) \\
286 & $PrBr$ & MPDS & s542061 & 166 & $R\overline{3}m$ & 0.0 & 0.0 & 0.0 & 0.0 \\
287 & $SmBi_{2}Se_{4}$ & MPDS & s543829 & 62 & $Pnma$ & -- & -- & -- & -- \\
288 & $PrSiI$ & MPDS & s1902103 & 164 & $P\overline{3}m1$ & 0.0 & 0.0 & 0.0 & 0.0 \\
289 & $NdIO$ & MPDS & s1140364 & 129 & $P4/nmm$ & 0.0 & 0.0 & 0.0 & 0.0 \\
290 & $KLaSiSe_{4}$ & MPDS & s537612 & 4 & $P2_1$(No.4) & 2.127 (I) & 2.072 (I) & 2.127 (I) & 2.072 (I) \\
291 & $TmSbS_{2}O$ & MPDS & s1237957 & 129 & $P4/nmm$ & 0.0 & 0.0 & 0.318 (D) & 0.166 (I) \\
292 & $LuBr_{3}$ & MPDS & s1300337 & 148 & $R\overline{3}$ & 3.605 (I) & 3.443 (I) & 3.669 (I) & 3.507 (I) \\
293 & $KDyI_{3}$ & MPDS & s1711765 & 63 & $Cmcm$ & 0.0 & 0.0 & 0.0 & 0.0 \\
294 & $HoBr_{3}$ & MPDS & s1300333 & 148 & $R\overline{3}$ & 0.0 & 0.0 & 3.517 (I) & 3.517 (I) \\
295 & $LaSbSe_{2}O$ & MPDS & s1502067 & 129 & $P4/nmm$ & 0.097 (I) & 0.0 & 0.097 (I) & 0.0 \\
\bottomrule
\end{longtable}